# Java File Security System (JFSS)

By Brijender Kahanwal, Kanishak Dua & Girish Pal Singh

*Maharaja Ganga Singh University, Bikaner India*

*Abstract -* Nowadays, storage systems are increasingly subject to attacks. So the security system is quickly becoming mendatory feature of the data storage systems. For the security purpose we are always dependent on the cryptography techniques. These techniques take the performance costs for the complete system. So we have proposed the Java File Security System(JFSS). It is based on the on-demand computing system concept, because of the performance issues. It is a greate comback for the system performance. The concept is used because, we are not always in need the secure the files, but the selected one only.

In this paper, we have designed a file security system on Windows XP. When we use the operating system, we have to secure some important data. The date is always stored in the files, so we secure the important files well. To check the proposed functionality, we experiment the above said system on the Windows operating system. With these experiments, we have found that the proposed system is working properly, according to the needs of the users.

*GJCST-E Classification:* D.4.6

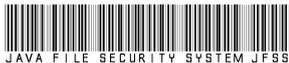

*Strictly as per the compliance and regulations of:*

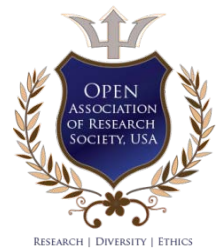

# Java File Security System (JFSS)

Brijender Kahanwal[α], Kanishak Dua[σ] & Girish Pal Singh[ρ]

*Abstract -* Nowadays, storage systems are increasingly subject to attacks. So the security system is quickly becoming mendatory feature of the data storage systems. For the security purpose we are always dependent on the cryptography techniques. These techniques take the performance costs for the complete system. So we have proposed the Java File Security System(JFSS). It is based on the on-demand computing system concept, because of the performance issues. It is a greate comback for the system performance. The concept is used because, we are not always in need the secure the files, but the selected one only.

In this paper, we have designed a file security system on Windows XP. When we use the operating system, we have to secure some important data. The date is always stored in the files, so we secure the important files well. To check the proposed functionality, we experiment the above said system on the Windows operating system. With these experiments, we have found that the proposed system is working properly, according to the needs of the users.

## I. Introduction

The access control is one of the fundamental security services in the computer system. It is a mechanism for constraining the interaction between users and protected resources. File is one of the important resources of the computer system. That must be protected from the unauthorized access that it can't be tempered or stolen by intruders. The file security can enforced using cryptographic techniques. With the help of these techniques the important files are encrypted and authorized users are given appropriate cryptographic keys.

The cryptographic techniques can be applied at any level of the storage systems because they use the layered architecture. The level may be the block or virtual one in the operating system. Basically, file management is an important task of the computer system. We have designed the Java File Security System (JFSS) [1-5] for files on the Windows XP.

The suggested file security system storing encrypted files using Rijndael Algorithm (AES) [6], so an unauthorized user can't access the important data. The encryption takes place for the selected files (important ones which requires the security) only.

We are using the concept of on-demand computing which results in the high performance of the computer system. The proposed system is working properly for all types of the files. In this paper there are more sections. Next section is section II which is about the related works. In section III, the design of the system is shown. In section IV, the evaluation is done. In section V, there is conclusion.

## II. Related Works

So many approaches are applied to solve the problem of information security. The approaches may be the user space or kernel space or the combined one. The kernel approach is sensitive to implement because any small mistake done by the programmer can harm the overall funtioning of the system. The user space one is secure and competible with the system and the independent one and comfortable in the implementation and are the highly portable if we are using the best portable platform like Java.

There are so many implementations in the literature review and every one has there advantages and disadvantages with them. BestCrypt [7], is designed as a loopback device driver which creates a raw block device with a single file. The single file acts as a container (the backing store). There is an associated cipher key for each container. Cryptographic File System (CFS) [8], provides a transparent UNIX file system interface to directory hierarchies that are automatically encrypted with user supplied keys. It is implemented as a user level NFS server. User needs to create an encrypted directory and assign its key which is required for cryptographic transformations, when the directory is created for the first time. Transparent Cryptographic File System (TCFS) [9], works as a layer under the Virtual File System (VFS) layer, making it completely transparent to the application. The security is applied by means of the Data Encryption Standard (DES) algorithm [10].

## III. Design

The main design goals of our research are as follows:
a) The proposed system should have better system performance as well as expand it for the existing file system.

*Author α :* Assistant Professor, CSE Department, GGGI, Dinarpur, Ambala, Haryana, INDIA. E-mail : imkahanwal@gmail.com
*Author σ :* Ex-B. Tech. Student, S. D. I. T. M., Israna, Panipat, Haryana, INDIA. E-mail : dua.kanishak@yahoo.co.in
*Author ρ :* Convener, CS and IT Department, Maharaja Ganga Singh University, Bikaner (Raj.), INDIA. E-mail : gpsbku@gmail.com

b) It should be independent of File System (it should not require the modifications in the other file systems or user applications).
c) It should offer strong storage security against most trivial and moderately sophisticated attacks.
d) It should be compatible with the future technology for separate key management just like smart cards for storing the encryption keys which are directly in the possession of authorized users.
e) It should be compatible with the existing file system services as the encrypted files should behave normally as of the other files within the system.
f) It should be developed as a user level file system and be convenient for users.

We have used the Windows XP operating system to design the functionality of file security system. The programming language to be used is the Sun Microsystems Java technology. To design it there is a function design form which has the necessary buttons on it.

The login form, that is used to login with the file security system. After entering the user id and password we are linking to the security execution program. We always need the user registration with the file security system. The registration is done by the program administrator who has the only permission to make number of users for the system. He or she will give the username and the password to the user. That is displayed in the Figure 3.1.

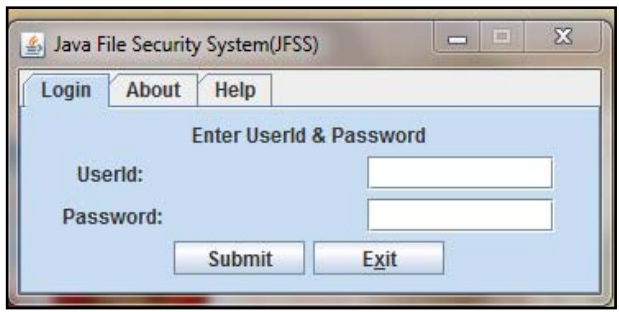

*Figure 3.1:* Login Display

After this user registration, he or she can login the system and use its functionality. The file encryption, decryption, about and help control form is appeared on the screen. It is shown in the Figure 3.2. It has the option to select the file to which the user wants to encrypt for the security feature. He or she can select any type of file and click on the encrypt button after that the encryption key is saved on the smart card is that is not available then the key is saved on the user specified location.

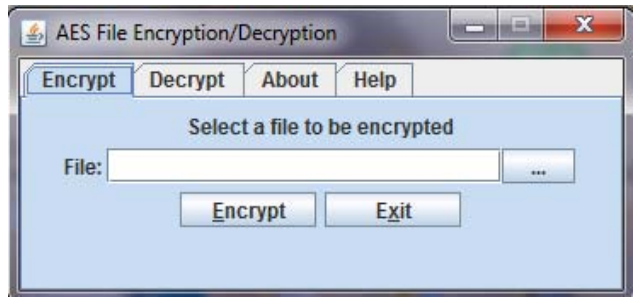

*Figure 3.2:* Encrytion tab display

The user may want to decrypt his previously encrypted file to use it. Then he or she have to make two selections one for the file and one for the key especialy the encryption key. Then the user will get the message to be successful or unsuccessful decryption. The successful message is shown in the Figure 3.3.

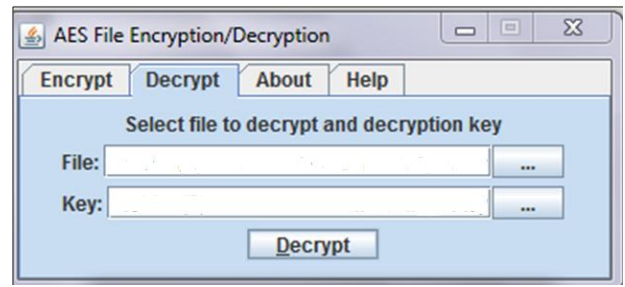

*Figure 3.3:* Decryption tab display

## IV. Evaluation

We performe test and evaluation on the proposed file security system for files and the directories. For experiment the computer system was with the configurations as Pentium 4 processor, Windows XP operating system.

The system has been tested for its functioning. In the first login window the user enter his or her userid and the password. If that is correct then he or she will get a message login successful or not. As in the Figure 4.1 the login is a successful one.

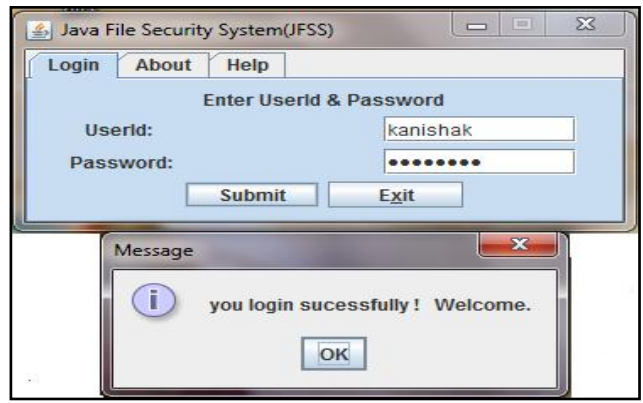

*Figure 4.1:* A successful login window

In the next screen shot the user is going to select an important file that has the need of security. It encryptes the specified file and save the encryption key to the smart card which is a sapeate location of storage from the encrypted file. It increses security of the data. It is shown in the Figure 4.2.

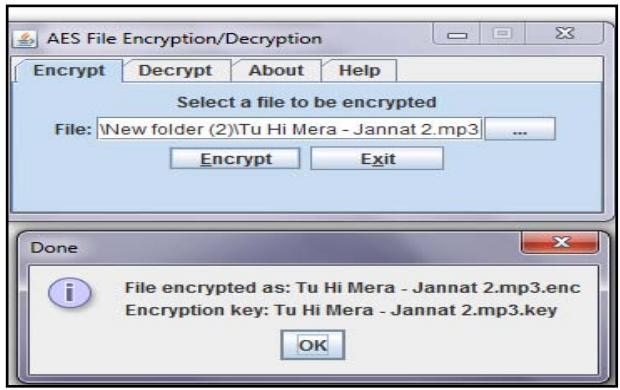

*Figure 4.2:* Encryption tab display

This is the Figure 4.3 which shows the decryption process of the system. It has two file selection buttons on it. One file selection button for the specified encrypted file to whom the user is going to decrypt. Another one is for the key selection of the specifed file. Because every file has its own independent key to encrypt or to decrypt it.

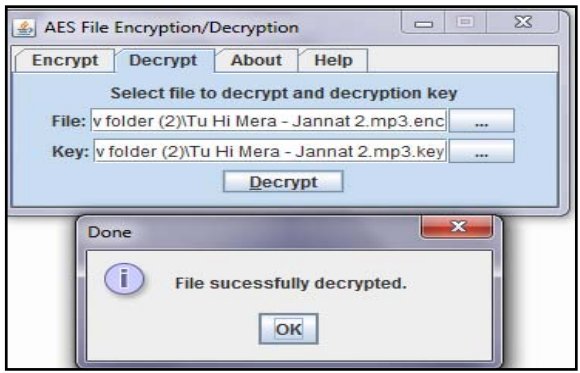

*Figure 4.2:* Decryption tab display

We have seen the file's look how it will behave after the encryption. The system is highly secure that we can cont delete the encrypted file and also con't change data which shows the integrity. The encrypted file's view is shown in the Figure 4.4.

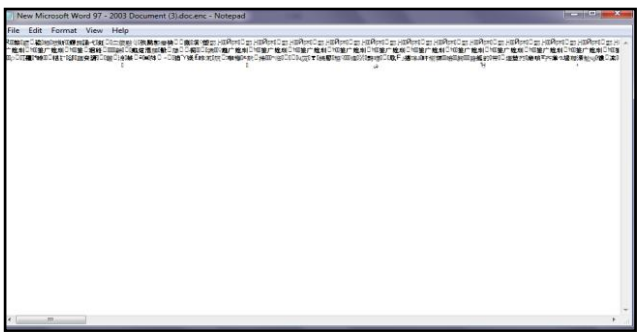

*Figure 4.4:* Display screen of an encrypted file

## V. Conclusion

We have contributed in the desiging and development of a user space cryptographic file system. We have balanced the design goals like security, performance, convenient and independability of the system.

We have achieved the high security by including the support of the Rijndeal Algorithm (AES) and we have saved the keys on the portable smart cards for the documents which are important.

The performance is achieved with the help of on-demand computing concept which is that we are not going to encrypt all the files on the computer system, but we are going to encyrpt only the important documents only. It saves the performance overhead of the system.

The system is very convenient to the users as described in the study done in the refernce [2]. And the independability is achieved with the help of the Java technology which is highly portable. So the complete system is a highly independent of the configuration.

## Acknowledgement

We would like to thank all the anonymous revieweres because of their valuable feedback and suggestions. We have developed a convenient system for the community.

## References Références Referencias

1. B. Kahanwal, T. P. Singh, and R. K. Tuteja. "A Windows Based Java File Security System (JFSS)". International Journal of Computer Science & Technology (2011), Vol. 2, No. 3, pp. 25-29.
2. B. Kahanwal, T. P. Singh, and R. K. Tuteja, "Java File Security System (JFSS) Evaluation Using Software Engineering Approaches", International Journal of Advanced Research in Computer Science & Software Engineering (2012), Vol. 2, No. 1, pp. 132-137.
3. B. Kahanwal, T. P. Singh, and R. K. Tuteja, "Towards the Framework of the File Systems Performance Evaluation Techniques and the Taxonomy of Replay Traces", International Journal of Advanced Research in Computer Science (2011), Vol. 2, No. 6, pp. 224-229.
4. B. Kahanwal, T. P. Singh, and R. K. Tuteja. "Performance Evaluation of Java File Security System (JFSS)", Pelagia Research Library—Advances in Applied Science Research (2011), Vol. 2, No. 6, pp. 254-260.
5. B. Kahanwal, and T. P. Singh, "Towards the Framework of Information Security", Journal of Current Engineering Research (2012), Vol. 2, No. 2, pp. 31-34.

6. Department of Commerce: National Institute of Standards and Technology (NIST), " Federal Information Processing Standard (FIPS) PUB #197: Advanced Encryption Standard (AES)", 2001, NIST, Gaithersuburg, MD, USA.
7. Mick Bauer, Paranoid penguin, "BestCrypt: Crossplatform filesystem Encryption", Linux Journal, 2002, 98:117.
8. M. Blaze, "A Cryptographic File System for UNIX", in ACM Conference on Computer and Communications Security,1993, pp. 9-16.
9. G. Cattaneo, L. Catuogno, A. D. Sorbo, and P. Persiono, "The Design and Implementation of a Transparent Cryptographic File System for UNIX", in the proceedings of USENIX Annual Technical Conference: FREENIX Track, 2001, pp. 245-252.
10. Department of Commerce: National Institute of Standards and Technology (NIST), " Federal Information Processing Standard (FIPS) PUB #81: Data Encryption Standard (DES)", NIST, Gaithersuburg, MD, USA, 1980.

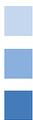